# Systematic study of deformed nuclei at the drip lines and beyond

M.V. Stoitsov,[1, 2, 3, 4] J. Dobaczewski,[5] W. Nazarewicz,[3, 4, 5] S. Pittel,[6] and D.J. Dean[3, 4]

[1]*Institute of Nuclear Research and Nuclear Energy,*
*Bulgarian Academy of Sciences, Sofia-1784, Bulgaria*
[2]*Joint Institute for Heavy Ion Research, Oak Ridge, Tennessee 37831*
[3]*Department of Physics and Astronomy, The University of Tennessee, Knoxville, Tennessee 37996*
[4]*Physics Division, Oak Ridge National Laboratory, P.O. Box 2008, Oak Ridge, Tennessee 37831*
[5]*Institute of Theoretical Physics, Warsaw University, Hoża 69, PL-00-681 Warsaw, Poland*
[6]*Bartol Research Institute, University of Delaware, Newark, Delaware 19716*

An improved prescription for choosing a transformed harmonic oscillator (THO) basis for use in configuration-space Hartree-Fock-Bogoliubov (HFB) calculations is presented. The new HFB+THO framework that follows accurately reproduces the results of coordinate-space HFB calculations for spherical nuclei, including those that are weakly bound. Furthermore, it is fully automated, facilitating its use in systematic investigations of large sets of nuclei throughout the periodic table. As a first application, we have carried out calculations using the Skyrme Force SLy4 and volume pairing, with exact particle number projection following application of the Lipkin-Nogami prescription. Calculations were performed for all even-even nuclei from the proton drip line to the neutron drip line having proton numbers $Z = 2, 4, \ldots, 108$ and neutron numbers $N = 2, 4, \ldots, 188$. We focus on nuclei near the neutron drip line and find that there exist numerous particle-bound even-even nuclei (i.e., nuclei with negative Fermi energies) that have at the same time negative two-neutron separation energies. This phenomenon, which was earlier noted for light nuclei, is attributed to bound shape isomers beyond the drip line.

## I. INTRODUCTION

The development of experimental facilities that accelerate radioactive ion beams [1, 2] has opened up a window to many nuclei that were heretofore inaccessible. With these new facilities and the new detector technology that is accompanying them, it is becoming possible to study the properties of nuclei very far from the valley of beta stability, all the way out to the particle "drip lines" and perhaps even beyond.

Much work is now in progress to develop appropriate theoretical tools for describing nuclei in these exotic regimes [3]. A proper theoretical description of such weakly-bound systems requires a careful treatment of the asymptotic part of the nucleonic density. An appropriate framework for these calculations is Hartree-Fock-Bogoliubov theory, solved in coordinate representation [4, 5, 6]. This method has been used extensively in the treatment of spherical systems but is much more difficult to implement for systems with deformed equilibrium shapes [7, 8, 9].

In the absence of reliable coordinate-space solutions to the deformed HFB equations, it is useful to consider instead the configuration-space approach, whereby the HFB solution is expanded in a single-particle basis. One approach has been to use a truncated basis composed partly of discrete localized states and partly of discretized continuum and oscillating states [7, 8, 10]. Because of the technical difficulties in implementing this method, it has typically been restricted to include states in the continuum up to at most several MeV. As a consequence, such an approach should not be able to describe adequately the spatial properties of nuclear densities at large distances.

An alternative possibility is to expand in a basis of spatially localized states. Expansion in a harmonic oscillator (HO) basis is particularly attractive because of the simple properties of oscillator states. There have been many configuration-space HFB+HO calculations reported, either employing Skyrme forces or the Gogny effective interaction [11, 12, 13, 14], or using a relativistic Lagrangian [15, 16]. This methodology has proven particularly useful when treating nuclei in or near the valley of stability. For nuclei at the drip lines, however, the HFB+HO expansion converges slowly as a function of the number of oscillator shells [6], producing wave functions that decrease too steeply at large distances. The resulting densities, especially in the pairing channel, are artificially reduced in the outer region and do not reflect correctly the pairing correlations of these weakly-bound nuclei.

A related approach that has recently been proposed is to instead expand the quasiparticle HFB wave functions in a complete set of transformed harmonic oscillator (THO) basis states [17, 18, 19], obtained by applying a local-scaling coordinate transformation (LST) [20, 21, 22] to the standard HO basis. The THO basis preserves many useful properties of the HO wave functions, including its simplicity in numerical algorithms, while at the same time permitting us to incorporate the appropriate asymptotic behavior of nuclear densities.

Applications of this new HFB+THO methodology have been reported both in the non-relativistic [18, 19] and relativistic domains [17]. In all of these calculations, specific global parameterizations were employed for the scalar LST function that defines the THO basis. There are several limitations in such an approach, however. On the one hand, any global parameterization of the LST function will of necessity modify properties throughout the entire nuclear volume, in order to improve the asymptotic density at large distances. This is not desirable, however, since the HFB+HO results are usually reliable in the nuclear interior, even for weakly-bound systems. In addition, because of the need to introduce matching conditions between the interior and exterior regions, a global LST function will invariably have a very complicated behavior, especially around the classical turning point, making it difficult to simply parameterize. Perhaps most importantly, the minimization procedure that is needed in such an approach to optimally define the basis parameters is computationally very time consuming, especially when a large number of shells is included, making it very difficult to apply the method systematically to nuclei across the periodic table.

In the present work, we propose a new prescription for choosing the THO basis. For a given nucleus, our new prescription requires as input the results from a relatively simple HFB+HO calculation, with no variational optimization. The resulting THO basis leads to HFB+THO results that almost exactly reproduce the coordinate-space HFB results for spherical [5] and axially deformed [10] nuclei and are of comparable quality to those of the former, more complex, HFB+THO methodology.

Because the new prescription requires no variational optimization of the LST function, it can be readily applied in systematic studies of nuclear properties. As the first such application, we carry out a detailed study of nuclei between the two-particle drip lines throughout the periodic table, using the Skyrme force SLy4 [23] and volume pairing [19]. In order to restore good particle number, we apply the Lipkin-Nogami (LN) prescription [24, 25, 26, 27, 28, 29] followed by exact particle-number projection (PNP) [30].

The structure of the paper is the following. In Sec. II, we briefly review the HFB and LN methods, noting several features particular to its coordinate and configurational representation. In Sec. III, we introduce the THO basis and then formulate our new prescription for the LST function. The results of systematic calculations of even-even nuclei are reported in Sec. IV, with special emphasis on those nuclei that are at the neutron drip line and just beyond. Conclusions and thoughts for the future are presented in Sec. V.

## II. OVERVIEW OF HARTREE-FOCK-BOGOLIUBOV THEORY AND THE LIPKIN-NOGAMI METHOD

In this section, we review the basic ingredients of Hartree-Fock-Bogoliubov theory and the Lipkin-Nogami method followed by particle-number projection. Since these are by now standard tools in nuclear structure, we keep the presentation brief and refer the reader to Ref. [30] for further details.

HFB is a variational theory that treats in a unified fashion mean-field and pairing correlations. The HFB equations can be written in matrix form as

$$\begin{pmatrix} h - \lambda & \Delta \\ -\Delta^* & -h^* + \lambda \end{pmatrix} \begin{pmatrix} U_n \\ V_n \end{pmatrix} = E_n \begin{pmatrix} U_n \\ V_n \end{pmatrix}, \quad (2.1)$$

where $E_n$ are the quasiparticle energies, $\lambda$ is the chemical potential, $h = t + \Gamma$ and $\Delta$ are the Hartree-Fock (HF) hamiltonian and the pairing potential, respectively, and $U_n$ and $V_n$ are the upper and lower components of the quasiparticle wave functions. These equations are solved subject to constraints on the average numbers of neutrons and protons in the system, which determine the two corresponding chemical potentials, $\lambda_n$ and $\lambda_p$.

In coordinate representation, the HFB approach consists of solving (2.1) as a set of integro-differential equations with respect to the amplitudes $U(E_n, \mathbf{r})$ and $V(E_n, \mathbf{r})$, both of which are functions of the position coordinate $\mathbf{r}$. The resulting density matrix and pairing tensor then read

$$\rho(\mathbf{r}, \mathbf{r}') = \sum_{0 \leq E_n \leq E_{\max}} V^*(E_n, \mathbf{r}) V(E_n, \mathbf{r}'), \quad (2.2a)$$

$$\kappa(\mathbf{r}, \mathbf{r}') = \sum_{0 \leq E_n \leq E_{\max}} V^*(E_n, \mathbf{r}) U(E_n, \mathbf{r}'). \quad (2.2b)$$

Typically, the HFB continuum is discretized in this approach by putting the system in a large box with appropriate boundary conditions [6].

In the configurational approach, the HFB equations are solved by matrix diagonalization within a chosen single-particle basis $\{\psi_\alpha\}$ with appropriate symmetry properties. In this sense, the amplitudes $U_n$ and $V_n$ entering Eq. (2.1) may be thought of as expansion coefficients for the quasiparticle states in the assumed basis. The nuclear characteristics of interest are determined from the density matrix and pairing tensor,

$$\rho(\mathbf{r}, \mathbf{r}') = \sum_{\alpha\beta} \rho_{\alpha\beta} \psi_\alpha(\mathbf{r}) \psi_\beta^*(\mathbf{r}'), \quad (2.3a)$$

$$\kappa(\mathbf{r}, \mathbf{r}') = \sum_{\alpha\beta} \kappa_{\alpha\beta} \psi_\alpha(\mathbf{r}) \psi_\beta(\mathbf{r}'), \quad (2.3b)$$

which are expressed in terms of the basis states $\psi_\alpha$ and the associated basis matrix elements as

$$\rho_{\alpha\beta} = \sum_{0 \leq E_n \leq E_{\max}} V_{\alpha n}^*(E_n) V_{\beta n}(E_n), \quad (2.4a)$$

$$\kappa_{\alpha\beta} = \sum_{0 \leq E_n \leq E_{\max}} V_{\alpha n}^*(E_n) U_{\beta n}(E_n). \quad (2.4b)$$

In configuration-space calculations, all quasiparticle states have discrete energies $E_n$.

The results from configuration-space HFB calculations should be identical to those from the coordinate-space approach when all the states $\psi_\alpha$ from a complete single-particle basis are taken into account. Of course, this is never possible. In the presence of truncation, it is essential that the basis produce rapid convergence, so that reliable results can be obtained within computational limitations on the number of basis states that can be included.

The LN method serves as an efficient method for restoring particle number before variation [24]. With only a slight modification of the HFB procedure outlined above, it is possible to obtain a very good approximation for the optimal HFB state, on which exact particle number projection then has to be performed [28, 31].

In more detail, the LN method is implemented by performing the HFB calculations with an additional term included in the HF hamiltonian,

$$h' = h - 2\lambda_2(1 - 2\rho), \quad (2.5)$$

and by iteratively calculating the constant $\lambda_2$ (separately for neutrons and protons) so as to properly describe the curvature of the total energy as function of particle number. For an arbitrary two-body interaction $\hat{V}$, $\lambda_2$ can be calculated from the particle-number dispersion according to [24],

$$\lambda_2 = \frac{\langle 0|\hat{V}|4\rangle\langle 4|\hat{N}^2|0\rangle}{\langle 0|\hat{N}^2|4\rangle\langle 4|\hat{N}^2|0\rangle}, \quad (2.6)$$

where $|0\rangle$ is the quasiparticle vacuum, $\hat{N}$ is the particle number operator, and $|4\rangle\langle 4|$ is the projection operator onto the 4–quasiparticle space. On evaluating all required matrix elements, one obtains [27]

$$\lambda_2 = \frac{4\text{Tr}\Gamma'\rho(1-\rho) + 4\text{Tr}\Delta'(1-\rho)\kappa}{8\left[\text{Tr}\rho(1-\rho)\right]^2 - 16\text{Tr}\rho^2(1-\rho)^2}, \quad (2.7)$$

where the potentials

$$\Gamma'_{\mu\mu'} = \sum_{\nu\nu'} V_{\mu\nu\mu'\nu'}(\rho(1-\rho))_{\nu'\nu}, \quad (2.8a)$$

$$\Delta'_{\mu\nu} = \tfrac{1}{2}\sum_{\mu'\nu'} V_{\mu\nu\mu'\nu'}(\rho\kappa)_{\mu'\nu'}, \quad (2.8b)$$

can be calculated in full analogy to $\Gamma$ and $\Delta$ by replacing the $\rho$ and $\kappa$ in terms of which they are defined by $\rho(1-\rho)$ and $\rho\kappa$, respectively. In the case of the seniority pairing interaction with strength $G$, Eq. (2.7) simplifies to

$$\lambda_2 = \frac{G}{4}\frac{\text{Tr}(1-\rho)\kappa\,\text{Tr}\rho\kappa - 2\,\text{Tr}(1-\rho)^2\rho^2}{[\text{Tr}\rho(1-\rho)]^2 - 2\,\text{Tr}\rho^2(1-\rho)^2}. \quad (2.9)$$

An explicit calculation of $\lambda_2$ from Eq. (2.7) requires calculating new sets of fields Eq. (2.8), which is rather cumbersome. However, we have found [32] that Eq. (2.7) can be well approximated by the seniority-pairing expression Eq. (2.9) with the effective strength

$$G = G_{\text{eff}} = -\frac{\bar{\Delta}^2}{E_{\text{pair}}} \quad (2.10)$$

determined from the pairing energy

$$E_{\text{pair}} = -\frac{1}{2}\text{Tr}\Delta\kappa \quad (2.11)$$

and the average pairing gap

$$\bar{\Delta} = \frac{\text{Tr}\Delta\rho}{\text{Tr}\rho}. \quad (2.12)$$

The use of the LN method in HFB theory requires special consideration of the asymptotic properties of quasiparticle states [4, 5], of essential importance for weakly-bound systems. Because of the modified HF hamiltonian (2.5), new terms appear in the HFB+LN equation, which are non-local in coordinate representation and thus can modify the asymptotic conditions. Effectively, this means that the standard Fermi energy $\lambda$ has to be replaced by

$$\lambda' = \lambda + 2\lambda_2(1 - 2n_{\min}) \quad (2.13)$$

or by

$$\lambda'' = \lambda + 2\lambda_2, \quad (2.14)$$

where $n_{\min}$ is the norm of the lower HFB component $V(E_{\min}, \boldsymbol{r})$ corresponding to the smallest quasiparticle energy $E_{\min}$.

The first expression (2.13) assumes that the asymptotic properties can be inferred from the HFB equation in the canonical basis, in which $\rho$ is diagonal and has eigenvalues that can be estimated by norms of the second HFB components. The second expression (2.14) pertains to the HFB equation in coordinate representation, in which the integral kernel $\rho(\boldsymbol{r}, \boldsymbol{r}')$ vanishes at large distances. Neither of these expressions can be rigorously justified, thereby demonstrating limitations of using the LN method to analyze spatial properties of wave functions. These ambiguities are enhanced by the fact that the LN method overestimates the curvature $\lambda_2$ near magic numbers [28, 31].

Note that in the exact projection before variation method, the Fermi energy is entirely irrelevant, and hence one should not attribute too much importance to the choice between $\lambda'$ and $\lambda''$. Nevertheless, since the PNP affects only occupation numbers, leaving the canonical wave functions unchanged, in what follows we use the modified Fermi energy $\lambda'$ in modelling the asymptotic behavior needed to implement the THO method.

Finally, we should note that the HFB machinery detailed above can be readily implemented with a quadrupole constraint [30], as is the case for some of the calculations we will be reporting.

### III. THE TRANSFORMED HARMONIC OSCILLATOR BASIS

In the present study, we carry out HFB calculations in configuration space, expanding in a transformed harmonic oscillator basis. This basis was originally introduced in Refs. [17, 18, 19], and we refer the reader to

Ref. [19] for details concerning the use of the deformed THO basis and for a discussion of the cut-off procedure that is used to perform the summations in Eq. (2.4). We also refer the reader to an interesting new application of the THO basis to one-dimensional problems of interest in molecular physics [33].

As noted earlier, all previous calculations using the THO basis in HFB calculations employed a global parameterization of the LST function that defined the basis. In the following subsections, we develop a new and improved form for the transformation, which we then use in the HFB+THO applications to be reported in Sect. V.

### A. Comparison of coordinate-space HFB calculations and configuration-space HFB+HO calculations

The main differences between the results of coordinate-space HFB calculations and those from configuration-space HFB+HO calculations can be seen in plots of the corresponding local density distributions. A typical example is shown in Fig. 1, where the densities and their logarithmic derivatives from coordinate-space HFB calculations (solid lines) are compared with those from a configurational HFB+HO calculation. Although the calculations were done for a specific spherical nucleus and Skyrme interaction, the features exhibited are generic. Note that the coordinate-space HFB calculations were carried out in a box of 30 fm, so that the logarithmic derivative of the density obtained in that calculation shows a sudden drop near the box edge.

Invariably, the logarithmic derivative $\rho'/\rho$ associated with the coordinate-space HFB solution shows a well-defined minimum near some point $R_{\min}$ in the asymptotic region, after which it smoothly approaches a constant value $-k$, where

$$k = 2\kappa = 2\sqrt{2m(E_{\min} - \lambda')/\hbar^2} \quad (3.1)$$

is associated with the HFB asymptotic behavior for the lowest quasiparticle state that has the corresponding quasiparticle energy $E_{\min}$ (see Eq. (2.13) and Ref. [6]). This property is clearly seen in the upper panel of Fig. 1. One can also see that the HFB+HO densities and logarithmic derivatives are in almost perfect agreement with the coordinate-space results up to (or around) the distance $R_{\min}$. We conclude, therefore, that the HFB+HO densities are numerically reliable up to that point.

Moreover, the HFB value of the density decay constant $k=2\kappa$, when calculated from Eq. (3.1), is also correctly reproduced by the HFB+HO results. It is not possible to distinguish between the values of $k$ that emerge from the coordinate-space and harmonic-oscillator HFB calculations, both values being shown by the same line in the upper panel of Fig. 1.

Soon beyond the point $R_{\min}$, the HFB+HO density begins to deviate dramatically from that obtained in the coordinate-space calculation. For relatively small numbers of harmonic oscillator shells $N_{\rm sh}$, the logarithmic derivative of the HFB+HO density goes asymptotically to zero following the gaussian behavior of the harmonic oscillator basis. The resulting HFB+HO density does not develop a minimum around the point $R_{\min}$, as seen from the $N_{\rm sh} = 8$ curve shown in the upper panel of Fig. 1. When the number of harmonic oscillator shells $N_{\rm sh}$ increases, the HFB+HO solution tries to capture the correct density asymptotics. Due to the gaussian asymptotic of the basis, however, the logarithmic derivative of the HFB+HO density only develops oscillations around the exact solution (see the $N_{\rm sh} = 12$ and 20 curves in the upper panel of Fig. 1). As a result, the logarithmic derivative of the HFB+HO density is very close to the coordinate-space result around the mid point $R_m = (R_{\max} - R_{\min})/2$, where $R_{\max}$ is the position of the first maximum of the logarithmic derivative after $R_{\min}$.

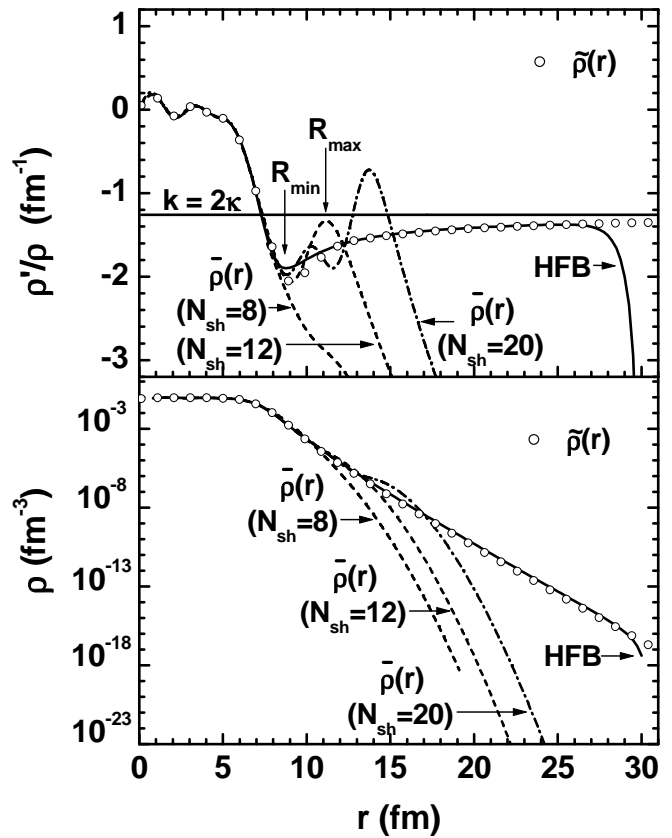

FIG. 1: Logarithmic derivative of the density (upper panel), and the density in logarithmic scale (lower panel), as functions of the radial distance. The coordinate-space HFB results (solid line) are compared with those for the HFB+HO method (denoted $\bar{\rho}$) with $N_{\rm sh}=$ 8, 12 and 20 HO shells, as well with the approximation (denoted $\tilde{\rho}$) given by Eq. (3.7) (small circles).

In summary, the following HFB+HO quantities agree with the coordinate-space HFB results: *(i)* the value of the density decay constant $k$; *(ii)* the local density up to the point $R_{\min}$ where the logarithmic derivative $\rho'/\rho$ shows a clearly-defined minimum; *(iii)* the actual value of this point $R_{\min}$; *(iv)* the value of the logarithmic derivative of the density at the point $R_m$ defined above. In fact, the last of the above is not established nearly as firmly as the first three; nevertheless, we shall make use of it in developing our new formulation of the HFB+THO method.

Beyond the point $R_m$, the HFB+HO solution fails to capture the physics of the coordinate-space results, especially in the far asymptotic region. It is this incorrect large-$r$ behavior that we now try to cure by introducing the THO basis.

### B. Approximation to the coordinate-space HFB local densities

Our goal is to try to find an approximation to the *exact* (coordinate-space) HFB density that is based only on information contained in the HFB+HO results. Towards that end, we make use of the WKB asymptotic solution of the single-particle Schrödinger equation for a given potential $V(r)$, assuming that beyond the classical turning point only the state with the lowest decay constant $k=2\kappa$ contributes to the local density. Under this assumption, the logarithmic derivative of the density can be written as

$$\left.\frac{\rho'(r)}{\rho(r)}\right|_{r\longrightarrow\infty} = -\frac{2}{r} - 2\sqrt{\kappa^2+\mathcal{V}} - \frac{1}{2}\frac{\mathcal{V}'}{\kappa^2+\mathcal{V}}, \quad (3.2)$$

where the first term comes from the three-dimensional volume element, while the next two correspond to the first- and second-order WKB solutions [34]. The reduced potential $\mathcal{V}$,

$$\mathcal{V}(r) = \frac{2m}{\hbar^2}V(r) = \mathcal{V}_N + \frac{\ell(\ell+1)}{r^2} + \frac{2m}{\hbar^2}\frac{Ze^2}{r}, \quad (3.3)$$

is the sum of the nuclear, centrifugal, and Coulomb (for protons) contributions, with $\ell$ being the single-particle orbital angular momentum.

In practical applications, it turns out that near $R_m$ the next-to-lowest quasiparticle states still contribute to the local density $\rho$ in a way that may be more important than the second-order WKB term shown in Eq. (3.2). Moreover, in deformed nuclei the quasiparticle states do not have good total angular momentum $\ell$, so that several quasiparticles may contribute to the asymptotic density depending on their $\ell$-content and the value of $\kappa$. Therefore, we need a practical prescription to fix a reasonable approximate asymptotic form of the density with minimal numerical effort but high reliability. This can be achieved by using in (3.2) a reduced potential of the form

$$\mathcal{V}(r) = \frac{C}{r^2} + \frac{2m}{\hbar^2}\frac{Ze^2}{r}, \quad (3.4)$$

where the nuclear part $\mathcal{V}_N$ (which is small around and beyond $R_m$) is neglected, and the coefficient $C$ is allowed to differ from its centrifugal barrier value $\ell(\ell+1)$. The actual value of $C$ is fixed by the requirement that the logarithmic derivative (3.2) coincides at the mid point $R_m$ with the $\ell$=0 component of the HFB+HO density, i.e., with

$$\bar{\rho}(r) = \int_0^{\pi/2} \bar{\rho}(r,\theta) P_{\ell=0}(\cos(\theta))\sin(\theta)d\theta. \quad (3.5)$$

Next, in order to make a smooth transition from the HFB+HO density $\bar{\rho}(r)$ in the inner region to the approximate asymptotic expression (3.2) in the outer region, we introduce the following approximation $\tilde{\rho}$ for the logarithmic density derivative:

$$\frac{\tilde{\rho}'(r)}{\tilde{\rho}(r)} = \begin{cases} \frac{\bar{\rho}'(r)}{\bar{\rho}(r)} & \text{for } r \leq R_{\min}, \\ a\frac{(R_{\min}-r)^2}{r^s} + b & \text{for } R_{\min} \leq r \leq R_{\max}, \\ -\frac{2}{r} - 2\sqrt{\kappa^2+\mathcal{V}} - \frac{1}{2}\frac{\mathcal{V}'}{\kappa^2+\mathcal{V}} & \text{for } r \geq R_{\max}. \end{cases} \quad (3.6)$$

The coefficients $a$ and $b$, and the power $s$, are determined from the condition that the logarithmic derivative (3.6) and its first derivative are smooth functions at the points $R_{\min}$ and $R_{\max}$. Note that the first derivative of (3.6) at $R_{\min}$ is automatically equal to zero, so that there is no need to introduce a fourth parameter to satisfy this condition.

Having determined the smooth expression for the logarithmic derivative of $\tilde{\rho}(r)$, we can derive the approximate local density distribution $\tilde{\rho}(r)$ by simply integrating Eq. (3.6). The result is

$$\tilde{\rho}(r) = \begin{cases} \bar{\rho}(r) & \text{for } r \leq R_{\min}, \\ A\,e^{-br}\exp\left[-\frac{a}{r^s}\left(\frac{ar^3}{3-s} - \frac{2r^2R_{\min}}{2-s} + \frac{rR_{\min}^2}{1-s}\right)\right] & \\ & \text{for } R_{\min} \leq r \leq R_{\max}, \\ B\,\frac{\exp\left[-2\int^r\sqrt{\kappa^2+\mathcal{V}}dr\right]}{r^2\sqrt{\kappa^2+\mathcal{V}}} & \text{for } r \geq R_{\max}, \end{cases} \quad (3.7)$$

where the integration constants $A$ and $B$ are determined from the matching conditions for the density at points $R_{\min}$ and $R_{\max}$, respectively. Finally, $\tilde{\rho}(r)$ is normalized to the appropriate particle number.

The approximate density (3.7) works fairly well for all nuclei that we have considered. This is illustrated in Fig. 1 where the approximate density $\tilde{\rho}$ (circles) is seen to be in perfect agreement with the coordinate-space HFB results.

It should be stressed that the above procedure is applicable only when the number of shells is large enough that the HFB+HO density has a minimum at the point $R_{\min}$. The minimum value of $N_{\text{sh}}$ required to satisfy this condition depends on the particular deformations or on the nuclei considered. For the number of shells $N_{\text{sh}} = 20$ used in our calculations, the above condition is always satisfied.

## C. LST function for HFB+THO calculations

The starting point of our new and improved HFB+THO procedure is, thus, to carry out a standard HFB+HO calculation for the nucleus of interest, thereby generating its local density and its local $\ell=0$ density $\bar{\rho}(r)$ (3.5), and then to use the method outlined in the previous subsection to correct that density at large distances (see Eq. (3.7)), by calculating $\tilde{\rho}(r)$. The next step is to define the LST [19] so that it transforms the HFB+HO $\ell=0$ density (3.5) into the corrected density of Eq. (3.7). This requirement leads to the following first-order differential equation,

$$\tilde{\rho}(r) = \frac{f^2(\mathcal{R})}{\mathcal{R}^2} \frac{\partial f(\mathcal{R})}{\partial \mathcal{R}} \bar{\rho}\left(\frac{r}{\mathcal{R}} f(\mathcal{R})\right) , \qquad (3.8)$$

which for the initial condition $f(0) = 0$ can always be solved for $f(\mathcal{R})$.

Once the LST function has been so obtained, we need simply diagonalize the HFB matrices in the corresponding THO basis. Most importantly, no information is required to build the THO basis beyond the results of a standard HFB+HO calculation. Since no further parameters enter, there is no need to minimize the HFB+THO total energy. As a consequence, with this new methodology we are able to systematically treat large sets of nuclei within a single calculation.

Despite the fact that the new HFB+THO method is simpler to implement than the earlier version, there are no discernible differences between the results obtained with the two distinct treatments of the LST function. Most importantly, the current formulation leads to the same excellent reproduction of coordinate-space results as did the previous one [18, 19].

## IV. RESULTS

In this section, we present the results of calculations performed for all particle-bound even-even nuclei with $Z\leq 108$ and $N\leq 188$. The THO basis was implemented according to the prescription developed in the previous section. The $k$ value used in the procedure was obtained in the following way. From the starting HFB+HO calculation, we determined $k$ values separately for neutrons and protons, using Eq. (3.1). We then associated the $k$ value for the transformation with the smaller of $k_p$ and $k_n$. In this way, the THO basis is always adapted to the less-bound type of particle. The calculations were performed by building THO basis states from spherical HO bases with $N_{\rm sh}=20$ HO shells and with oscillator frequencies of $\hbar\omega_0 = 1.2 \times 41 \,{\rm MeV}/A^{1/3}$.

In order to meaningfully test predictions of nuclear masses for neutron-rich nuclei, we used the SLy4 Skyrme force parameterization [23], as this was adjusted with special emphasis on the properties of neutron matter. At present, there also exist Skyrme forces that were adjusted exclusively to nuclear masses [35]. These forces were used within a calculation scheme that was not focused on weakly-bound nuclei. In the pairing channel, we used a pure volume contact pairing force $V^\delta(\boldsymbol{r},\boldsymbol{r}')=V_0\delta(\boldsymbol{r}-\boldsymbol{r}')$ with strength $V_0 = -167.35\,{\rm MeV\,fm}^3$ and acting within a phase space limited by a cut-off parameter [19] of $\bar{e}_{\rm max} = 60\,{\rm MeV}$.

Figure 2 summarizes the systematic results of our calculations, both for ground state quadrupole deformations (upper panel) and for two-neutron separation energies (lower panel). For this figure, calculations for a given mass number $A$ were carried out for increasing (decreasing) $N-Z$, up to the nucleus with positive neutron (proton) Fermi energy. Furthermore, for each nuclide, three independent sets of HFB+THO+LN calculations were performed, for initial wave functions corresponding to oblate, spherical, and prolate shapes, respectively. Depending on properties of a given nucleus, we could therefore obtain one, two, or three solutions with different shapes. For each obtained solution we performed a PNP calculation of the total energy. The lowest of these energies for a given nucleus was then identified with the ground-state solution.

Calculations of a microscopic mass table are greatly helped by taking advantage of parallel computing. We have used two IBM-SP computers at ORNL: Eagle, a 1 Tflop machine, and Cheetah, a 4 Tflop machine (1 Tflop = $1 \times 10^{12}$ operations/second). The code performs at 350 Mflop/processor on Eagle. We created a simple load-balancing routine that allows us to scale the problem to 200 processors. We are able to calculate the entire deformed even-even mass table in a single 24 wall-clock hour run (or approximately 4,800 processor hours). A complete calculated mass table is available online in Ref. [36].

The ground-state quadrupole deformations $\beta$ displayed in Fig. 2 (upper panel) were estimated from the HFB+THO+LN total quadrupole moments and rms radii through a simple first-order expression [30]. In that panel, all even-even nuclei with negative Fermi energies, $\lambda_n<0$ and $\lambda_p<0$, are shown. In the lower panel, showing two-neutron separation energies $S_{2n}$, results are shown for those $N$ and $Z$ values for which the nuclides with both $N$ and $N-2$ have $\lambda_n<0$. Note that on the proton-rich side the lighter of them may have $\lambda_p>0$; nevertheless, we show these points to make the proton drip line in the $S_{2n}$ panel identical to that of the quadrupole deformation panel. Of course, on the proton drip line values of $S_{2n}$ are large and not very illuminating.

Table I summarizes our results for even-even nuclei along the two-particle drip lines. More specifically, for each value of $Z$, the results for the lightest isotope with $\lambda_p<0$, and the heaviest isotope with $\lambda_n<0$ are presented.

As can be seen from Fig. 2 and Table I, our calculations produce several particle-bound even-even nuclei (i.e., nuclei with negative Fermi energies) that at the same time have negative two-proton (or two-neutron) separation energies. Such an effect was already noticed in light nuclei in Ref. [19]. The current calculations suggest it may be generic, occurring near both the two-neutron and two-proton drip lines and for nuclei as light as $^{42}$Mg and as heavy as $^{216}$Dy. It seems to be related to the fact that the Fermi energies pertain to stability with respect to particle emission of a given configuration or shape, namely that of the ground state. In many of the cases in which we observe this phenomenon, (a) the neighboring even-even nucleus, the one to which it would decay by two-nucleon emission, has two distinct shapes, each with negative Fermi energies, (b) the ground state of that neighboring nucleus has a shape that is different than that of the parent nucleus, (c) the shape of the excited bound configuration is the same as that of the parent nucleus, and (d) decay to the excited configuration is energetically forbidden.

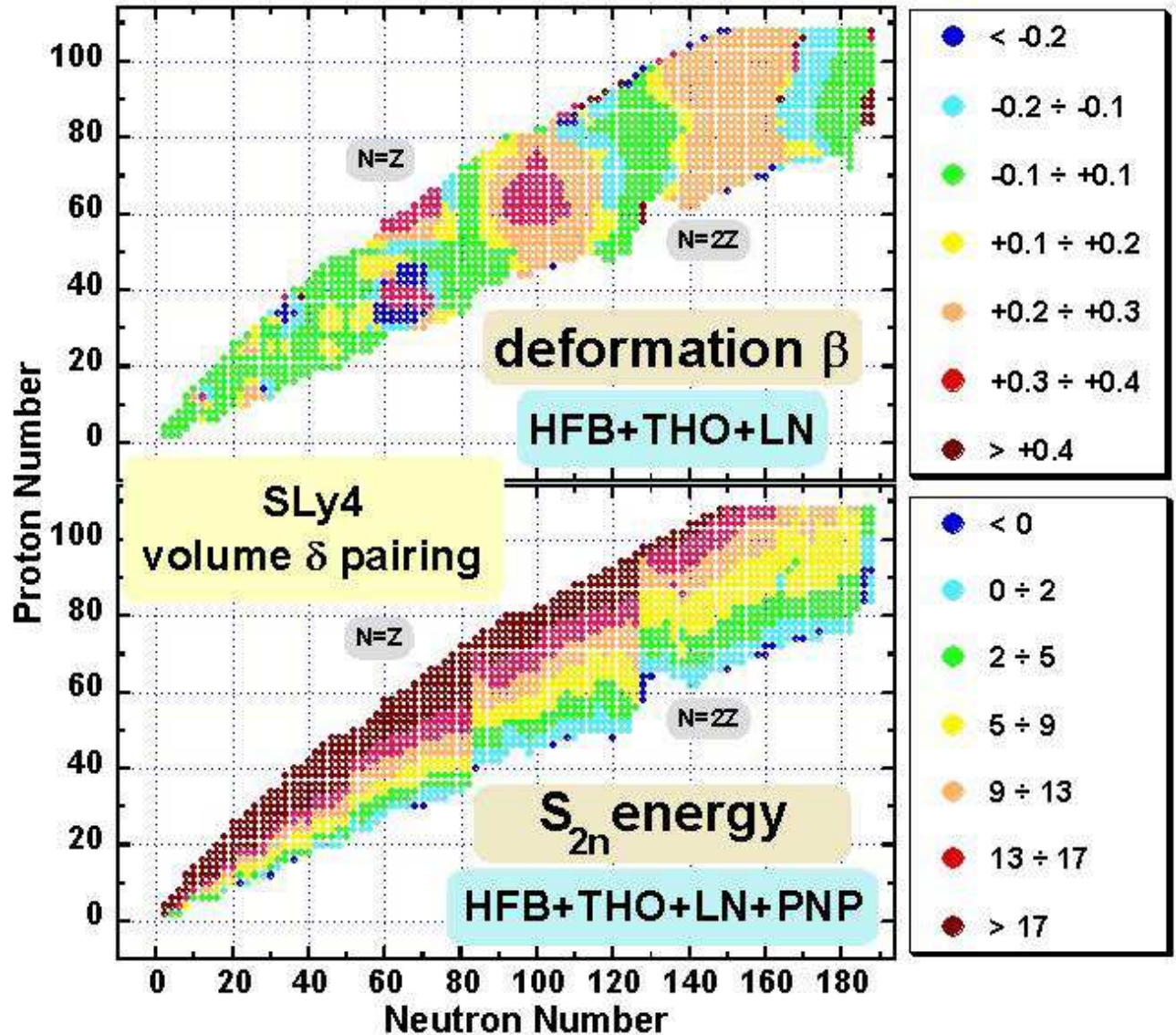

FIG. 2: Quadrupole deformations $\beta$ (upper panel) and two-neutron separation energies $S_{2n}$ in MeV (lower panel) of particle-bound even-even nuclei calculated within the HFB+THO method with Lipkin-Nogami correction followed by exact particle number projection. The Skyrme SLy4 interaction and volume contact pairing were used.

The precise results of course depend sensitively on properties of the interaction, both in the particle-hole and particle-particle channels. Despite its many good features, the force we use is far from perfect. For example, the positions we obtain for the two-neutron drip lines in the Be and O isotopes are not correct. In addition, the method itself has limitations, as it leaves out potentially important effects beyond mean field. Despite these limitations, we feel it is nevertheless worthwhile to point out some of the interesting new physical situations that are predicted in these calculations and which may therefore occur in weakly-bound systems. The above example of nuclei that are formally beyond one of the two-particle drip lines but nevertheless are localized and do not spontaneously spill off a nucleon is just one of several. We will now discuss in greater detail some specific isotopic chains to see how this and other interesting exotic new features emerge.

TABLE I: Results of the HFB+THO calculations for drip-line nuclei with the SLy4 Skyrme force and volume delta pairing force. The left and right columns show results for proton and neutron drip-line isotopes from He to Pb. For both drip lines we show deformations $\beta$, Fermi energies $\lambda$ (in MeV), two-particle separation energies (in MeV), and neutron and proton pairing gaps (in MeV).

| | Two-proton drip line | | | | | | Two-neutron drip line | | | | |
|---|---|---|---|---|---|---|---|---|---|---|---|
| Nucleus | $\beta$ | $\lambda_p$ | $S_{2p}$ | $\bar{\Delta}_n + \lambda_{2n}$ | $\bar{\Delta}_p + \lambda_{2p}$ | Nucleus | $\beta$ | $\lambda_n$ | $S_{2n}$ | $\bar{\Delta}_n + \lambda_{2n}$ | $\bar{\Delta}_p + \lambda_{2p}$ |
| $^4$He | 0.00 | $-10.49$ | | 5.59 | 5.50 | $^8$He | 0.00 | $-1.26$ | 2.69 | 2.71 | 5.35 |
| $^6$Be | 0.00 | $-2.13$ | 1.79 | 5.51 | 2.89 | $^{12}$Be | 0.00 | $-2.70$ | 6.92 | 2.70 | 2.76 |
| $^{10}$C | 0.00 | $-4.38$ | 11.44 | 3.03 | 3.15 | $^{22}$C | 0.00 | $-0.34$ | 2.97 | 2.03 | 2.69 |
| $^{14}$O | 0.00 | $-3.76$ | 10.80 | 3.17 | 2.86 | $^{26}$O | 0.00 | $-0.97$ | 0.53 | 1.53 | 2.87 |
| $^{18}$Ne | 0.00 | $-3.46$ | 7.26 | 2.96 | 1.85 | $^{34}$Ne | 0.28 | $-0.39$ | 0.50 | 1.40 | 1.76 |
| $^{20}$Mg | 0.00 | $-1.64$ | 2.76 | 2.98 | 1.84 | $^{42}$Mg | $-0.18$ | $-0.29$ | $-0.44$ | 1.09 | 1.64 |
| $^{24}$Si | $-0.07$ | $-2.65$ | 5.63 | 1.85 | 1.87 | $^{46}$Si | 0.00 | $-0.99$ | 1.71 | 1.07 | 1.86 |
| $^{28}$S | 0.00 | $-2.08$ | 6.10 | 1.92 | 1.92 | $^{52}$S | 0.00 | $-0.05$ | $-0.96$ | 1.00 | 1.49 |
| $^{32}$Ar | 0.00 | $-1.85$ | 4.50 | 2.15 | 1.48 | $^{58}$Ar | 0.00 | $-0.39$ | 2.37 | 1.31 | 1.39 |
| $^{36}$Ca | 0.00 | $-1.49$ | 5.24 | 1.77 | 1.76 | $^{68}$Ca | 0.00 | $-0.11$ | 0.40 | 1.10 | 1.73 |
| $^{40}$Ti | 0.00 | $-0.95$ | 2.31 | 1.74 | 1.26 | $^{72}$Ti | 0.00 | $-0.63$ | 2.59 | 1.15 | 1.05 |
| $^{44}$Cr | 0.00 | $-1.57$ | 3.58 | 1.94 | 1.30 | $^{80}$Cr | $-0.00$ | $-0.07$ | 0.01 | 0.72 | 1.14 |
| $^{46}$Fe | 0.00 | $-0.25$ | 1.07 | 1.94 | 1.31 | $^{84}$Fe | 0.00 | $-0.12$ | 0.60 | 0.80 | 1.15 |
| $^{52}$Ni | $-0.03$ | $-1.45$ | 3.74 | 1.37 | 1.56 | $^{88}$Ni | 0.00 | $-0.19$ | 0.09 | 0.91 | 1.53 |
| $^{56}$Zn | 0.13 | $-0.57$ | 2.45 | 1.39 | 1.24 | $^{100}$Zn | 0.24 | $-0.02$ | $-0.29$ | 0.90 | 1.10 |
| $^{60}$Ge | $-0.09$ | $-0.17$ | 0.63 | 1.67 | 1.22 | $^{108}$Ge | 0.16 | $-0.13$ | 0.12 | 0.93 | 1.07 |
| $^{64}$Se | $-0.17$ | $-0.15$ | 0.83 | 1.25 | 1.27 | $^{114}$Se | 0.08 | $-0.27$ | 0.69 | 0.91 | 1.08 |
| $^{70}$Kr | $-0.22$ | $-1.10$ | 2.67 | 1.38 | 1.10 | $^{118}$Kr | 0.00 | $-0.23$ | 3.29 | 1.20 | 1.08 |
| $^{72}$Sr | 0.36 | $-0.16$ | $-1.74$ | 1.26 | 1.18 | $^{120}$Sr | 0.00 | $-0.86$ | 4.61 | 1.23 | 1.06 |
| $^{76}$Zr | 0.00 | $-0.19$ | 0.89 | 1.37 | 1.25 | $^{124}$Zr | 0.00 | $-0.04$ | $-0.74$ | 0.60 | 1.05 |
| $^{82}$Mo | 0.00 | $-0.83$ | 2.09 | 1.37 | 0.98 | $^{132}$Mo | 0.00 | $-0.05$ | 0.14 | 0.66 | 0.87 |
| $^{86}$Ru | 0.00 | $-0.83$ | 2.27 | 1.13 | 0.98 | $^{142}$Ru | 0.27 | $-0.02$ | 0.23 | 0.84 | 0.89 |
| $^{90}$Pd | 0.07 | $-0.90$ | 2.57 | 1.11 | 0.93 | $^{150}$Pd | $-0.22$ | $-0.02$ | $-0.44$ | 0.84 | 0.82 |
| $^{94}$Cd | 0.00 | $-0.88$ | 1.72 | 1.08 | 0.89 | $^{168}$Cd | $-0.02$ | $-0.01$ | $-0.62$ | 0.82 | 0.75 |
| $^{102}$Sn | 0.00 | $-0.80$ | 6.03 | 0.99 | 1.54 | $^{174}$Sn | 0.00 | $-0.27$ | 1.11 | 0.76 | 1.16 |
| $^{108}$Te | 0.16 | $-1.00$ | 2.39 | 1.13 | 0.89 | $^{176}$Te | 0.00 | $-0.83$ | 1.90 | 0.78 | 0.77 |
| $^{112}$Xe | 0.22 | $-0.83$ | 2.54 | 1.10 | 0.88 | $^{178}$Xe | 0.00 | $-1.37$ | 2.82 | 0.80 | 0.83 |
| $^{116}$Ba | 0.32 | $-1.02$ | 2.60 | 1.07 | 0.87 | $^{182}$Ba | 0.00 | $-0.28$ | 4.36 | 1.26 | 0.87 |
| $^{118}$Ce | 0.37 | $-0.19$ | 1.71 | 1.12 | 0.87 | $^{186}$Ce | 0.43 | $-0.11$ | $-16.29$ | 0.72 | 0.88 |
| $^{124}$Nd | 0.38 | $-0.33$ | 1.98 | 0.98 | 0.93 | $^{188}$Nd | 0.44 | $-0.51$ | $-15.32$ | 0.75 | 0.71 |
| $^{130}$Sm | 0.36 | $-0.64$ | 2.09 | 1.00 | 0.83 | $^{204}$Sm | 0.28 | $-0.01$ | 0.11 | 0.69 | 0.75 |
| $^{134}$Gd | 0.36 | $-0.44$ | 1.60 | 0.99 | 0.82 | $^{208}$Gd | 0.29 | $-0.20$ | 0.84 | 0.73 | 0.74 |
| $^{138}$Dy | 0.36 | $-0.12$ | 0.78 | 0.98 | 0.82 | $^{216}$Dy | $-0.22$ | $-0.02$ | $-4.70$ | 0.73 | 0.71 |
| $^{144}$Er | $-0.19$ | $-0.41$ | 1.64 | 0.89 | 0.89 | $^{222}$Er | 0.28 | $-0.08$ | 0.16 | 0.65 | 0.70 |
| $^{148}$Yb | $-0.16$ | $-0.11$ | 0.85 | 0.88 | 0.86 | $^{230}$Yb | $-0.21$ | $-0.00$ | $-0.06$ | 0.70 | 0.71 |
| $^{152}$Hf | $-0.10$ | $-0.05$ | 0.59 | 0.82 | 0.92 | $^{254}$Hf | 0.00 | $-0.02$ | | 0.72 | 0.86 |
| $^{158}$W | $-0.06$ | $-0.50$ | 1.36 | 0.84 | 0.94 | $^{256}$W | 0.00 | $-0.30$ | | 0.70 | 0.83 |
| $^{162}$Os | 0.11 | $-0.09$ | 0.57 | 0.84 | 0.78 | $^{258}$Os | 0.00 | $-0.57$ | 0.51 | 0.67 | 0.79 |
| $^{168}$Pt | 0.14 | $-0.04$ | 0.43 | 0.96 | 0.66 | $^{260}$Pt | 0.00 | $-0.83$ | 1.19 | 0.65 | 0.73 |
| $^{172}$Hg | $-0.08$ | $-0.04$ | $-1.13$ | 1.14 | 0.69 | $^{262}$Hg | 0.00 | $-1.09$ | 2.37 | 0.62 | 0.69 |
| $^{182}$Pb | 0.00 | $-0.11$ | 1.65 | 1.24 | 1.38 | $^{266}$Pb | 0.00 | $-0.03$ | 3.21 | 1.06 | 0.98 |

We focus our discussion on the heaviest isotopes of four isotopic chains; neon, magnesium, sulfur, and zinc (see Figs. 3–6, respectively). The figures show the Fermi energies $\lambda_n$, $\lambda'_n$, and $\lambda''_n$ [see Eqs. (2.13) and (2.14)], and the total binding energies, obtained in constrained

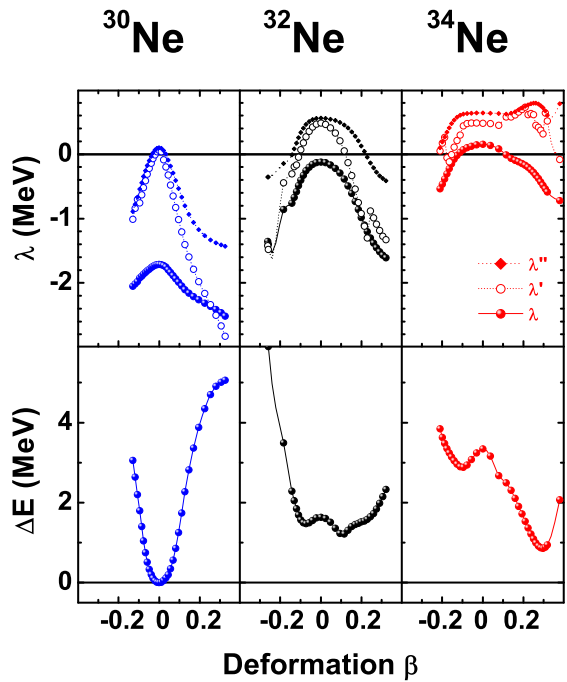

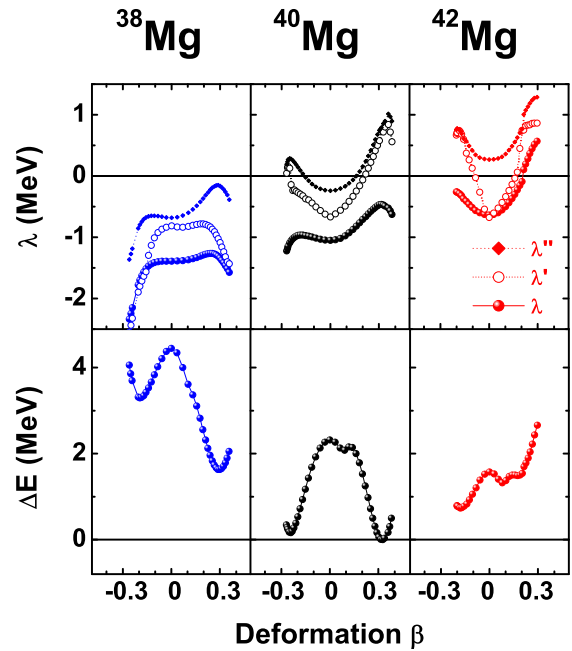

FIG. 3: Neutron Fermi energies $\lambda$ (upper panels) and the total binding energies (lower panels) calculated for $^{30}$Ne, $^{32}$Ne, and $^{34}$Ne as functions of the quadrupole deformation $\beta$.

FIG. 4: Same as in Fig. 3 but for $^{38}$Mg, $^{40}$Mg, and $^{42}$Mg.

HFB+THO+LN+PNP calculations as functions of the quadrupole deformation $\beta$ for the last three particle-bound isotopes of the respective chains. In each figure, the binding energies of the last three isotopes are shown on a common energy scale. As a reminder, two neutron separation energies can be readily obtained from the binding energies according to $S_{2n} = E(Z, N-2) - E(Z, N)$.

We should note that the minima of the constrained energies need not exactly correspond to the PNP of the HFB+THO+LN minima, which were used in Fig. 2 and Table I. Indeed, in the constrained calculations the deformation serves as an additional variational parameter for the variation *after* PNP. Optimally, the full variation after projection should be performed, which, however, requires a much larger numerical effort, and is left to future work. Such an optimal method will also remove the ambiguities related to the definition of the Fermi energy, discussed in Sec. II. At present, we illustrate these ambiguities by showing in Figs. 3–6 the three possible values of the Fermi energy, $\lambda_n$, $\lambda'_n$, and $\lambda''_n$.

Consider first the Ne isotopes, for which the results are shown in Fig. 3. For the SLy4 interaction that we use, a strong shell gap at $N=20$ persists up to the heaviest isotopes of Ne, and this produces a stiff spherical minimum for $^{30}$Ne. Adding two neutrons gives rise to the nucleus $^{32}$Ne, which is particle-bound ($\lambda_n<0$), but at the same time two-neutron unstable ($S_{2n}<0$). [Note that this nucleus does not exactly fit into the picture given earlier for such nuclei.] Interestingly, when we add two more neutrons, we obtain a strongly (prolate) deformed particle-bound ground configuration in $^{34}$Ne, which is again two-neutron stable ($S_{2n}>0$).

Next we turn to the Mg isotopes, for which results are presented in Fig. 4. In $^{40}$Mg the neutron Fermi energies $\lambda_n$ have negative values for all deformations, so that the configurations for all deformations are particle-bound, with the prolate minimum being slightly lowest. The same is also true for the next nucleus $^{42}$Mg where the ground state deformation changes from prolate to oblate. It is clear from Fig. 4 that in $^{42}$Mg the two-neutron separation energy is negative; however, since $^{42}$Mg and $^{40}$Mg have different shapes in their ground states, the real process of emitting two neutrons may occur towards the shape isomer in $^{40}$Mg. (The situation will be even more complicated if the oblate minimum in $^{42}$Mg is unstable to triaxial deformations, i.e., it is a saddle point.)

The results for the S isotopes are given in Fig. 5. Here, the spherical HFB+THO+LN minimum in $^{52}$S is shifted in the constrained PNP calculations towards a small oblate deformation. All shapes appear to be very weakly particle-bound, and have negative two-neutron separation energies at the same time. It is obvious that in the case of so poorly defined a minimum, its precise location is not relevant and full configuration mixing, e.g., within the generator coordinate method (GCM) [30, 37, 38], should be applied. This complication specific to weakly-bound nuclei is related to the fact that it is not clear how to take into account in the GCM the regions

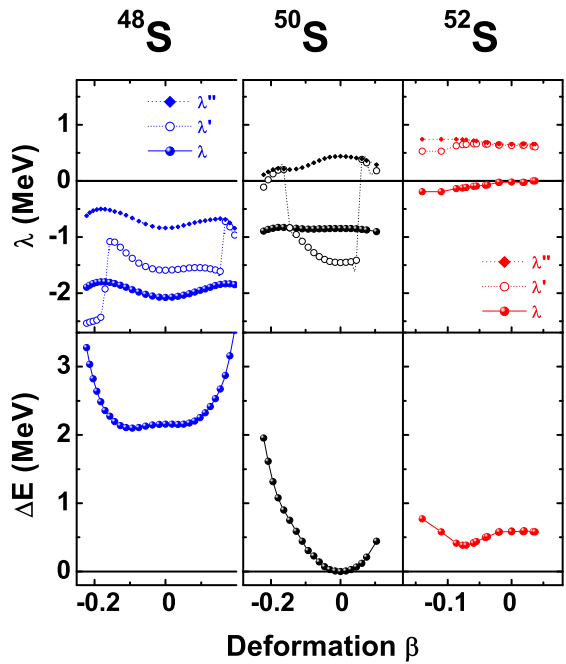

FIG. 5: Same as in Fig. 3 but for $^{48}$S, $^{50}$S, and $^{52}$S.

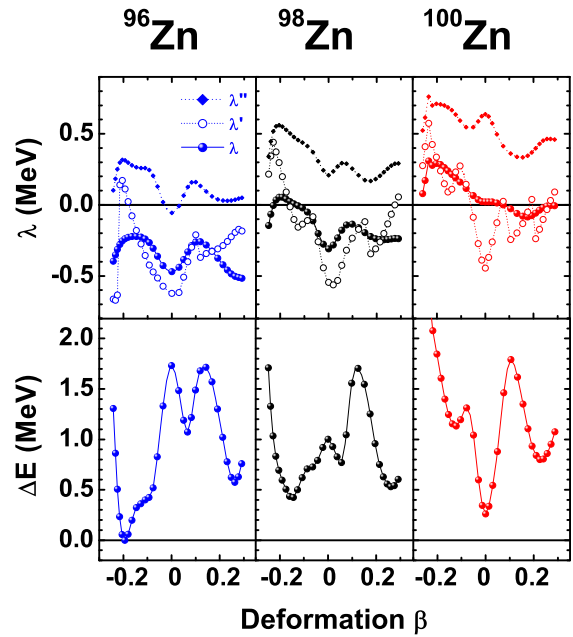

FIG. 6: Same as in Fig. 3 but for $^{96}$Zn, $^{98}$Zn, and $^{100}$Zn.

of the collective coordinate corresponding to $\lambda>0$, hence to particle-unbound states.

In the results for the zinc isotopes (Fig. 6), we see strong competition between oblate, prolate, and spherical shapes. In $^{96}$Zn, all shapes are particle-bound and the ground state is oblate. The situation changes in $^{98}$Zn, where the oblate configuration, though lowest in energy, becomes particle-unbound and the prolate minimum becomes the ground-state configuration. Though this ground state is two-neutron unstable ($S_{2n}<0$), its decay to the ground state of $^{96}$Zn may be hindered by the shape change. Finally, in $^{100}$Zn the particle-stable prolate ground state is also two-neutron unstable. Hence in this isotopic chain the last *two* even isotopes are unstable with respect to two-neutron emission.

In heavier nuclei near the neutron drip line, we often obtain particle-stable and two-neutron-unstable isotopes right after closed neutron magic shells. As in Ne, this reflects the fact that strong shell gaps persist up to the heaviest isotopes in a chain when the calculations are based on the SLy4 interaction. In the $N$=126 isotopes of Ce and Nd, for example, the ground-state configurations are strongly spherical. In the neighboring $N$=128 isotopes, these spherical configurations become particle unbound. However, in these same isotopes, there are strongly prolate particle-bound configurations with very large negative two-neutron separation energies (see Table I). An analogous situation occurs in the $N$=186 and 188 drip-line nuclei, where the last *two* even isotopes may have particle-bound prolate states with unbound spherical configurations.

Strong SLy4 neutron magic numbers also result in the characteristic non-monotonic behavior of the $S_{2n}$ values (Fig. 2). Indeed, lines of constant $S_{2n}$ often follow decreasing $Z$ with increasing $N$, which is particularly conspicuous near $N$=126. This effect even creates a small *peninsula* of stability near $N$=140. Such strong neutron closed shells could create the well-known deficiencies in the r-process abundances [39].

## V. CONCLUDING REMARKS

In this paper, we have reported the development of an improved version of the configuration-space HFB method expanded in a transformed harmonic oscillator basis. In its current form, the method can be used reliably in systematic studies of wide ranges of nuclei, both spherical and axially deformed, extending all the way out to the nucleon drip lines. The key step was the development of a prescription for choosing a reliable transformation function to define the THO basis that does not require variational optimization. The current prescription only involves information from a preliminary configuration-space HFB calculation carried out in a harmonic oscillator basis. The transformation function is then tailored to correct the asymptotic properties of the HFB+HO results. The resulting HFB+THO theory accurately reproduces results of coordinate-space HFB theory, where available, and also reproduces the results obtained with an earlier version of the transformation that had to be optimized separately for each nucleus.

As a first application of the new HFB+THO methodology, we carried out a systematic study of all even-even nuclei having $Z \leq 108$ and $N \leq 188$. Variation after particle-number projection was approximately included using the Lipkin-Nogami method, with exact projection performed for the final self-consistent solutions. We focussed our discussion on those nuclei that are very near the nucleon drip lines, finding that in several regions of the periodic table there exist nuclei that are stable against one-particle emission but unstable against pair emission. We showed that invariably this is associated with a shape change in the ground state. For example, while two-particle emission to the configuration of the daughter with the same shape as the parent is forbidden, a decay to the ground state having a different shape can nevertheless occur. The associated change in shape may conceivably lead to sufficient hindrance of the decay, hence the longer lifetime. Consequently, it is conceivable that there exist nuclei that formally live beyond the neutron drip line but can be observed experimentally. This phenomenon, which had earlier been noted in calculations of light nuclei, is now seen to be a more common feature of nuclei near the neutron drip line.

In the description of very weakly-bound systems, small changes in the results can have important consequences, changing for example the precise locations of the drip lines. It is important, therefore, to continue to improve the current HFB+THO methodology to accommodate effects not presently being included. Particularly important could be effects that arise beyond mean field. It is also important to develop the new HFB+THO formalism for application to odd-mass systems, including the effects of Pauli blocking. But most crucial, in our opinion, is to develop new-generation energy density functionals that will allow for more reliable predictions of the properties of exotic nuclei. Work along these various lines is currently underway.

## VI.  ACKNOWLEDGMENTS

This work has been supported in part by the Bulgarian National Foundation for Scientific Research under project Φ-809, by the Polish Committee for Scientific Research (KBN) under Contract No. 5 P03B 014 21, by the Foundation for Polish Science (FNP), by the U.S. Department of Energy under Contract Nos. DE-FG02-96ER40963 (University of Tennessee), and DE-AC05-00OR22725 with UT-Battelle, LLC (Oak Ridge National Laboratory), by the US National Science Foundation under grant Nos. PHY-9970749 and PHY-0140036, and by the computational grants from the *Regionales Hochschulrechenzentrum Kaiserslautern* (RHRK), Germany, and from the Interdisciplinary Centre for Mathematical and Computational Modeling (ICM) of the Warsaw University.